\documentclass[12pt]{article}
\usepackage{geometry}                
\usepackage{times,psfrag,graphicx,theorem,epsfig,lscape,amssymb,amsmath,dcolumn,
multirow,epic,float,enumerate,latexsym,ifthen,pifont,comment,bm,subfigure,color}
\usepackage[round]{natbib}
\usepackage{rotating}
\usepackage{url}
\usepackage[normalem]{ulem}
\newcommand{\blinding}[2]{#1}   

\renewcommand{\baselinestretch}{1.60}

\theoremstyle{plain} 
\theoremstyle{plain} 

\DeclareGraphicsExtensions{.png, .pdf, .jpg, .mps}
\theoremstyle{plain} \newtheorem{ass}{Assumption}

\DeclareMathOperator\bE{\mathbb E} 

\newcommand{\one}{\mathbf{1}}

\newcommand{\bx}{\mathbf{x}}
\newcommand{\bX}{\mathbf{X}}

\begin{document}

\begin{center}

{\Large Do debit cards decrease cash demand? Evidence from a causal analysis using Principal Stratification}

\medskip
\blinding{
Andrea Mercatanti \footnote{Economic and Financial Statistics Department, Bank of Italy, Rome, Italy. Email: mercatan@libero.it} \quad \quad  Fan Li \footnote{Department of Statistical Science, Duke University, Durham, NC, USA. Email: fli@stat.duke.edu  \\ Mercatanti's research was partially supported by the U.S. National Science Foundation (NSF) under Grant DMS-1127914 to the Statistical and Applied Mathematical Sciences Institute (SAMSI). Li's research was partially funded by NSF-SES grants 1155697 and 1424688. The content is solely the responsibility of the authors and does not necessarily represent the official views of Bank of Italy, SAMSI or NSF.}

}{}

\end{center}

{\centerline{ABSTRACT} \noindent It has been argued that innovation in transaction technology may modify the cash holding behaviour of agents, as debit card holders may either withdraw cash from ATMs or purchase items using POS devices at retailers. In this paper, within the Rubin Causal Model, we investigate the causal effects of the use of debit cards on the cash inventories held by households using data from the Italy Survey of Household Income and Wealth (SHIW). We adopt the principal stratification approach to incorporate the share of debit card holders who do not use this payment instrument. We use a regression model with the propensity score as the single predictor to adjust for the imbalance in observed covariates. We further develop a sensitivity analysis approach to assess the sensitivity of the proposed model to violation to the key unconfoundedness assumption. Our empirical results suggest statistically significant negative effects of debit cards on the household cash level in Italy.

\noindent

\vspace*{0.5cm}
\noindent {\sc Key words}: causal inference, potential outcomes, principal stratification, propensity score, sensitivity, unconfoundedness.
}

\clearpage

\section{Introduction}
Since the early 1970s, the diffusion of payment cards, such as debit, credit, pre-paid cards, raises the concern whether our societies are transforming into cashless societies. Cash has not disappeared in the meanwhile, however the interest in the question remains, particularly for what concerns the wide diffusion of debit cards. Debit cards are defined as cards enabling the holder to have purchases directly charged to funds on his account at a deposit-banking institution \citep{CPSS01}. The fact that debit cards allow withdraws at ATM points justifies the need of less cash holdings; at the same time the possibility to pay directly at POS devices at retailers makes this payment instrument a close substitute for cash. Moreover, both of the operations, withdrawal and payment, involve very low, or most of the times null, costs for debit cards compared to the rest of noncash instruments.

The possible effect of the use of debit cards on cash holdings has also important implications for central banks as the sole issuer of cash. First, substitution of cash for cards could lead to decrease in seignorage incomes. Because banknotes represent non-interest-bearing central bank liabilities, early concerns emerged among policymakers fearing that the cash shrinkage would have lead to a decline in central bank asset holdings and, consequently, in seigniorage revenue \citep{Stix04}. Second, in low interest rate regimes, the cash-card substitution can be very sensitive to interest rate, and consequently interest rate adjustments as a monetary policy to influence bank lending may be difficult \citep{Markose03, Yilmazkuday09}. The existing literature has been focused on the impact that payment cards may have on the functional relationship of the cash demand to other variables that central banks use to implement monetary policy \citep{Duca95, Attanasio02, LippiSecchi09, AlvarezLippi09} while relatively little attention has been directed to quantification of the effects of debit cards on the level of cash inventories held by individuals or households.

This paper aims to evaluate the causal effect of debit cards on cash inventories using data from the Survey on Household Income and Wealth (SHIW), a bi-annually national survey run by Bank of Italy on several aspects of Italian household economic and financial behaviour. We adopt the Rubin Causal Model (RCM) \citep{Rubin74, Rubin78} to conduct the causal analysis. Under the RCM, for each post-treatment variable, each unit has a potential outcome corresponding to each treatment level, and a causal effect on that post-treatment variable is defined as a comparison between the corresponding potential outcomes on a common set of units. A critical requirement of the RCM is that a ``cause'' (or a treatment without distinction) must or at least conceptually be manipulatable -- the principle of ``no causation without manipulation'' \citep{Holland86}. This raises a conceptual challenge in our application because using debit cards is largely a voluntary activity, and it is not clear that we could expose a person to use debit cards in any verifiable sense. Instead it is more natural to conceive the possession of cards---a status controlled by banks that issue cards---as the treatment variable. However, the primary research interest obviously lies in the effect of using cards rather than possessing cards. In fact, a significant portion of Italian households in the SHIW who have debit cards do not use them. Moreover, even among card users, there is various degree of usage: some use debit cards only to withdraw cash from ATM occasionally, while others use them frequently for both cash withdraw and payment at retailers. The issue is not limited to Italy, for example data from a survey conducted in Austria in 2003 reports evident shares of non-users and different use frequencies among debit cards holders \citep{Stix04}. Naturally there may be heterogeneous effects of debit cards on cash holding among different groups of card users.

We propose to address these complications via principal stratification \citep{FrangakisRubin02}, a unified framework for causal inference in the presence of post-treatment variables. The key is to treat the possession of debit cards as the treatment and the use of cards as a post-treatment intermediate variable between the treatment and the outcome. Principal stratification is a cross-classification of the population based on the joint potential values of an intermediate variable under each of the treatment, i.e., principal strata, and the interest lies in estimating causal effects local to certain principal strata. For example, the focus in our application is the causal effect in the stratum of units who possess and use debit cards -- the ``compliers''. This is similar to the instrumental variable approach to noncompliance in randomized experiments \citep{Angrist96, ImbensRubin97, Hirano00}, a special case of principal stratification. Recently a rapid growing literature has extended principal stratification to a wide range of settings in both experimental and observational studies, including ``censoring by death'' \citep{Rubin06, MatteiMealli07, Zhang09}, missing data \citep{Mattei14}, surrogate endpoints \citep{Gilbert03, LiTaylorElliott09, LiTaylorElliott11}, mediation analysis \citep{Gallop09, Elliott10}, and designs \citep{MatteiMealli11}. 
More complex settings such as ordinal or continuous intermediate variables have also been explored \citep{Frangakis04, JinRubin08, Griffin08, Schwartz11}.

Because only one of the two potential outcomes is observed for each unit, principal strata are latent, and thus identification of principal causal effects relies on assumptions such as unconfoundedness and exclusion restrictions. Unconfoundedness is particularly crucial in our study given the observational nature of SHIW. We tackle this issue in two steps. First, we propose a model-based regression approach where, to reduce the risk of mis-specification due to the imbalance in the multiple covariates between treatment groups, we use the estimated propensity score \citep{RosenbaumRubin83a} as the sole predictor. Then, we design and conduct a comprehensive sensitivity analysis on unconfoundedness: we encode the degree of violation to unconfoundedness as sensitivity parameters in the assumed regression model and examine how estimates change over plausible range of the sensitivity parameters. This vein of sensitivity analysis in causal inference originates from \cite{RosenbaumRubin83b}. In particular, our method is built upon that of \cite{Schwartz12}, who elaborated the various pathways of confounding and examined the sensitivity to both unconfoundedness and exclusion restriction, in the context of principal stratification. Alternative sensitivity analysis approaches have also been developed in the literature \citep[e.g.][]{Grilli08, Sjolander09, Jo11, StuartJo13, Gilbert03}.

The rest of the article is organized as follows. In Section 2, we introduce the basic setup, estimand and assumptions under principal stratification. In Section 3, we propose a model-based approach for estimation and a sensitivity analysis under the model. Section 4 presents the data and the empirical results. Section 5 concludes.

\section{Basic setup under principal stratification}
\subsection{Notions and estimands}
Because debit cards are typically issued to individuals, the natural statistical units would be individuals possessing debit cards in our analysis. However, SHIW collects information only on the household level. To mitigate the problem, we set household as the unit, but limit the sample of treated units to the households possessing one and only one debit card during the study period. This ensures a possible effect on household cash holding will be due only to the individual possessing the card, who is usually the head of the household.

Consider the study sample consists of $N$ units. For unit $i$, let $Z_i$ be the binary treatment, equal to 1 if the household possesses one and only one debit card and 0 otherwise; $D_i$ be the binary post-treatment variable, equal to 1 if the household uses a debit card and 0 otherwise\footnote{We will define two different characterization for the use of debit cards in Section \ref{sec:data}.};  $Y_i$ be the outcome, the average amount of cash held by the household;  and $\bX_i$ be the set of pre-treatment covariates. Because utilization of debit cards is a post-treatment event, we can define its corresponding potential outcomes: let $D_i(1)$ and $D_i(0)$ be the potential debit card usage status if unit $i$ does and does not have a card, respectively, equal 1 if the unit uses the card and 0 otherwise. Similarly, let $Y_i(1)$ and $Y_i(0)$ be the potential outcome, if unit $i$ does and does not possess a debit card, respectively. These notations of potential outcomes imply the acceptance of the Stable Unit Treatment Value Assumption (SUTVA; Rubin, 1980), that is, no interference between the units and no different versions of a treatment. SUTVA is deemed reasonable in this study, because the holding of debit cards in one household does not seem to affect the potential debit card utilization or cash inventory of other households. For each unit $i$, we only observe the potential outcomes corresponding to the observed treatment: $D_i=D_i(Z_i)$, $Y_i=Y_i(Z_i)$.

A principal stratification with respect to a post-treatment intermediate variable $D$ is a partition of units based on the joint potential values of $D$, i.e., principal strata: $S_i=(D_i(0), D_i(1))$. When both $Z$ and $D$ are binary, there are four principal strata in theory: $S_i \in \{(0,0), (0,1), (1,0), (1,1)\}$. In this application, obviously one can not use debit cards without possessing one, and also there are units who possess cards but not use them. As such, our study sample automatically satisfies a strong `monotonicity' condition:
\begin{ass} \label{monotonicity} (Monotonicity).
(1) $D_i(0)=0$; (2) $0<\Pr(D_i = 0|\bX_i,Z_i=1)<1$, for all $i$.
\end{ass}
Under monotonicity, there are only two principal strata: $S_i=(0,0)=n$, units who would not use debit cards irrespective of whether possessing one, and $S_i=(0,1)=c$, units who would use debit cards if possessing one but would not use if otherwise. We refer to these two strata as never-users and compliers, respectively, following the nomenclature of noncompliance in \cite{Angrist96}.  

By definition the principal stratum membership $S_i$ is not affected by the treatment assignment. Therefore, comparisons of $Y(1)$ and $Y(0)$ within a principal stratum are well-defined causal effects because they compare quantities defined on a common set of units. Here our interest lies in estimating the causal effects for the treated compliers, that is, units possessing and using debit cards. Thus we define the targeted estimand to be the average causal effect of the treated compliers (CATT):
\begin{equation}
\text{CATT}\equiv \bE[Y_i(1)-Y_i(0) \mid S_i=c, Z_i=1]=\bE_{\bx|Z=1}\{\bE[Y_i(1)-Y_i(0) \mid S_i=c, Z_i=1, \bX_i=\bx]\}.
\label{CATT}
\end{equation}
Analogously we can define the compliers average treatment effect (CATE), also known as the local average treatment effect (LATE) \cite{ImbensAngrist94}:
\begin{equation}
\text{CATE}\equiv \bE[Y_i(1)-Y_i(0) \mid S_i=c] =\bE_{\bx}\{\bE[Y_i(1)-Y_i(0) \mid S_i=c, \bX_i=\bx]\}.
\label{CATE}
\end{equation}
Both CATE and CATT are intention-to-treat (ITT) effects, representing effects of possessing debit cards, rather than effects of using cards. To attribute these effects to the use of cards, we make the following exclusion restriction assumption for the compliers, following the established literature in the IV approach to noncompliance \citep[e.g.][]{Angrist96, ImbensRubin97}:
\begin{ass} \label{ER_compliers} (Exclusion Restriction for compliers).
For all units with $S_i=(0,1)$, the effect of card possession is only through using the card.
\end{ass}
A formalized version of this assumption, which requires double-indexed notations, is given in \citep{ImbensRubin15} (Assumption 23.4). This type of exclusion restriction is in fact routinely made, often implicitly, in randomized experiments with full compliance.

\subsection{Identification of the causal effects}
For the same unit, only one of the two potential outcomes $(D_i(0), D_i(1))$ is observed, and thus the principal stratum $S_i$ is at most partially observed. The following assumptions are necessary for establishing the nonparametric identifiability of CATT and CATE \citep{ImbensAngrist94}.

\begin{ass} \label{overlap} (Overlap).
$0<\Pr (Z_i=1|\bX_i)<1$, for all $i$.
\end{ass}

\begin{ass} \label{unconfoundedness} (Unconfoundedness). $\{Y_i(1), Y_i(0), D_i(1), D_i(0)\} \perp Z_i | \bX_i.$
\end{ass}

\begin{ass} \label{ER_nevertakers} (Exclusion Restriction for never-takers)
\[\bE[Y_i(1)|\bX_i, S_i=n]=\bE[Y_i(0)|\bX_i, S_i=n].\]
\end{ass}

Assumption \ref{overlap} is the standard overlap condition, stating that each unit has a positive probability of possessing a card.  Assumption \ref{unconfoundedness} states that the treatment assignment is independent of all the potential outcomes conditional on observed pre-treatment variables; it is also known as the assumption of ``no unmeasured confounders''. Assumption \ref{ER_nevertakers} states that possessing debit cards does not affect the outcome for never-takers. Though both are exclusion restrictions, Assumption \ref{ER_nevertakers} and Assumption \ref{ER_compliers} are of very different nature: the former is a necessary condition for identifying the causal effects, whereas the latter is made solely for interpreting the ITT effects as effects of the actual treatment received. More discussions on the difference can be found in \cite{MealliPacini13} and \cite{ImbensRubin15}(Chapter 23).

In randomized experiments, Assumptions \ref{overlap}-\ref{ER_nevertakers} automatically hold without conditioning on covariates, and CATT equals CATE. In observational studies, unconfoundedness generally relies on conditioning on a number of observed covariates, and CATT and CATE are usually different. Consequently analysts often adopt regression models to adjust for the covariates to estimate the causal effects. However, two complications often arise, as in this application: first, distributions of some covariates can be significantly imbalanced between treatment groups, leading a regression analysis to rely heavily on model specification \citep{Imbens04}; second, unconfoundedness may still be questionable even conditioning on a large number of observed covariates. To address the first issue, we propose to combine the propensity score method and regression adjustment.  To address the second issue, we conduct a comprehensive sensitivity analysis around the regression models.

\section{Model-based estimation and sensitivity analysis} \label{sec:models}
\subsection{Models}
In the context of principal stratification, six quantities are associated with each unit: $Y_i(0)$, $Y_i(1)$, $D_i(0)$, $D_i(1)$, $Z_i$, $X_i$. Under nonconfoundedness, the joint distribution of these quantities can be written as:
\begin{eqnarray}
&&\Pr(Y_i(0),Y_i(1), D_i(0),D_i(1), Z_i, X_i) \nonumber\\
&=& \Pr(Y_i(0),Y_i(1)|S_i, X_i)\Pr(S_i|X_i) \Pr(Z_i|X_i)\Pr(X_i) \nonumber\\
&=& \Pr(Y_i(0),Y_i(1)|S_i, e(X_i))\Pr(S_i|e(X_i))e(X_i)\Pr(X_i),
\label{eq: noconfound}
\end{eqnarray}
where $e(X_i)=\Pr(Z_i=1|X_i)$ is the propensity score. In the analysis, we will condition on the observed distribution of covariates, so that $\Pr(X_i)$ does not need to be modeled. Therefore, three models are required for inference: one for the propensity score, one for the principal strata given the propensity score, and one for the potential outcomes given principal stratum and the propensity score.

Compared to the nonparametric approach, the model-based approach is more flexible, can reduce bias and improve precision, and also offers conceptually straightforward ways to incorporate complexities like multilevel structure, multiple outcomes, and latent variables. However, parametric models on a large number of pre-treatment variables are also more sensitive to mis-specification \citep{Rubin79}. In particular, imbalance in the pre-treatment variables between treatment groups or between different strata renders causal effects estimation to rely heavily on extrapolation, and consequently, on the functional specification. Because the propensity score reduces the dimension from the space of covariates to one, and balance of the propensity score leads to balance of each observed covariate, an attractive alternative is to use the estimated propensity score as the sole covariate \citep[e.g.][]{Heckman98}. This method is not as efficient as the regression estimator
based on adjustment for all covariates when the model is correctly specified \citep{Hahn98}, and one has to estimate the propensity score, which is also subject to mis-specification. However, simulation studies in \cite{Mercatanti14a} show that, in the presence of covariate imbalance between treatment groups, mis-specification of the regression model leads much larger biases and mean square error (MSE) than mis-specification of the propensity score. Moreover, in a recent simulation-based study, \cite{Hade14} suggests that if the distributions of the estimated propensity score in the treated and untreated groups have different shapes but roughly the same support, as in our application, then regression on the estimated propensity score performs well compared to the conventional regression model on the entire set of covariates and other propensity score based methods (e.g. matching, stratification and weighting). Therefore, given the relatively large number of covariates in our application, we choose the regression on propensity score approach.


Since there are only two strata in our application, we use a logistic regression model for the principal stratum membership $S$:
\begin{eqnarray}
\mbox{logit}(\Pr(S_i=n|X_i=x)) &=& \alpha_0+e(x) \cdot \alpha. \label{eq:smodel}
\end{eqnarray}
And we assume a linear regression model for continuous potential outcomes, with different intercepts and slopes for different strata:
\begin{eqnarray}
\Pr(Y_i(z)| S_i, X_i=x)&=&\one_{S_i=c}\cdot (\beta_{c0} + z\cdot \theta_{c}+e(x)\cdot \beta_{c1}) +\one_{S_i=n} \cdot (\beta_{n0} +e(x)\cdot \beta_{n1}) + \epsilon_i, \nonumber\\
 \label{eq:ymodel}
\end{eqnarray}
where $\epsilon_i\sim N(0,\sigma^2)$ and $\one_{S_i=s}$ is an indicator function that equals one if
$S_i=s$ and equals zero otherwise. It is easy to show that $\theta_c$ is the CATE, and the CATT can be subsequently estimated by averaging the differences between the observed outcomes for treated compliers and their estimated counterfactuals:
\begin{eqnarray*}
\widehat{\mbox{CATT}} = \frac{\sum_{i}D_{i}\ \cdot \ Z_{i} \cdot [Y_{i}-(\hat{\beta}_{c0}+e(X_{i})\cdot \hat{\beta}_{c1})]}{\sum_{i}D_{i}\ \cdot \ Z_{i}}.
\end{eqnarray*}
The maximum likelihood (ML) estimates of the parameters are obtained using an EM (expectation-maximization) algorithm. In the E-step the unobserved principal strata are replaced by their expectations given the data and the current estimates of the parameters; then in the M-step, the likelihood conditional on the expected principal strata is maximized. Standard errors are obtained by the outer product of gradients evaluated at the ML estimate for the parameters in \eqref{eq:smodel} and \eqref{eq:ymodel}, and by the bootstrap for the CATT.


\subsection{Sensitivity analysis}
Our sensitivity analysis is in the same spirit of \cite{RosenbaumRubin83b}, where the assignment to treatment is assumed to be unconfounded given the observed covariates $X$ and an unobserved covariate $U$, but is confounded given only $X$. Rosenbaum and Rubin suggest to specify a set of parameters characterizing the distribution of $U$ and the association of $U$ with $Z$ and $Y(z)$ given observed covariates. Assuming a parametric model, the full likelihood for $Z, Y(0),Y(1), U$ given $X$ is derived and maximized, fixing the sensitivity parameters to a range of  known values, and the results are compared. In order to incorporate the additional complexity of the immediate variable in our setting, we adopt a simpler setup similar to that in \cite{Schwartz12}. In particular, we do not directly model the distributions involving $U$, instead we model the consequences of an unmeasured confounder. Specifically, in the presence of $U$, Equation \eqref{eq: noconfound} no longer stands, instead we have
\begin{eqnarray*}
\Pr(Y_i(0),Y_i(1),S_i|Z_i,X_i, U_i)
&=& \Pr(Y_i(0),Y_i(1)|Z_i,S_i,X_i,U_i) \Pr(S_i|Z_i,X_i,U_i). \label{eq:factorization}
\end{eqnarray*}
Therefore, even conditional on the observed covariates, the proportion of principal strata and the distribution of the potential outcomes within a principal stratum can differ across treatment groups. These two channels of confounding are referred to as  $S$-confounding and $Y$-confounding in \cite{Schwartz12}, and are encoded in the following two models modified from \eqref{eq:smodel} and \eqref{eq:ymodel}, respectively.

The principal strata model \eqref{eq:smodel} is expanded to account for the $S$-confounding:
\begin{eqnarray}
\mbox{logit}(\Pr(S_i=n|Z_i=z, X_i=x)) &=& \alpha_0+e(x)\cdot \alpha  + \xi\cdot z , \label{eq:smodel_confound}
\end{eqnarray}
where $\xi$ is the log odds ratio of being a never-taker among card-holders versus among non card-holders conditional on covariates:
\[\exp(\xi) = \frac{\Pr(S_i=n|Z_i=1, X_i)/\Pr(S_i=c|Z_i=1, X_i)}{\Pr(S_i=n|Z_i=0, X_i)/\Pr(S_i=c|Z_i=0, X_i)}.\]
The parameter $\xi$ is estimable from the observed data, but nevertheless can also be viewed as a sensitivity parameter: when the unconfoundedness assumption holds, the odds ratio should be 1 and $\xi=0$, and consequently large absolute value of the estimated $\xi$ suggests severe imbalance in the proportions of principal strata between treatment groups and thus large $S$-confounding. Here $\xi$ is imposed to be the same across different values of $X$. We have also fitted the model with an interaction term between $X$ and $Z$ but observed little difference in our application.

For the potential outcomes models, in the presence of unmeasured confounding, it is important to differentiate the assignment $z$ in the definition of potential outcomes and the observed assignment $Z$, and we make the distinction using different subscripts $z_1$ and $z_2$. Specifically, we expand model \eqref{eq:ymodel} by adding two sensitivity parameters as follows:
\begin{eqnarray}
&&\Pr(Y_i(z_1)|S_i, Z_i=z_2, X_i=x) \nonumber\\
&=&\one_{S_i=c}\cdot (\beta_{c0} + z_1\theta_{c}+ z_2\eta_c+e(x)\cdot \beta_c) +\one_{S_i=n} \cdot (\beta_{n0}+z_2 \eta_n+ e(x)\cdot \beta_n)  + \epsilon_i,
 \label{eq:ymodel_confound}
\end{eqnarray}
It is straightforward to show that $\eta_c$ and $\eta_n$ account for the $Y$-confounding for compliers and never-takers, respectively:
\begin{eqnarray*}
\eta_s = \Pr(Y_i(z_1)| Z_{2i}=1, S_i=s, X_i) - \Pr(Y_i(z_1)| Z_{2i}=0, S_i=s, X_i),
\end{eqnarray*}
for $s=n, c$. For parsimony, the model assumes that $\eta_c$ and $\eta_n$ are constant across $x$ and $z$. We have also conducted analysis with interaction terms between $X$ and $Z$, and the results are similar. In our application, $\eta_n$ (or $\eta_c$) is the difference in cash holding between a never-user (or complier) who has a debit card and a never-user (or complier) who does not have a card.
When the unconfoundedness assumption holds, $\eta_c=\eta_n=0$, and model \eqref{eq:ymodel_confound} reduces to model \eqref{eq:ymodel}.

The sensitivity analysis is carried out by comparing the ML estimates of the CATT while fixing the sensitivity parameters at a range of plausible values. Among the three parameters, $(\xi, \eta_c, \eta_n)$, $\xi$ is estimable from the data and we fix it at the estimated value. Since for each unit only the potential outcome corresponding to the observed treatment is observed, $z_1=z_2$ in all observed units in Model \eqref{eq:ymodel_confound}, and thus only the sum $\theta_c+\eta_c$ but not each individual parameter is identifiable. For example, one can not differentiate the two sets of parameters ($\theta_c, \eta_c$) and ($\theta_c + v, \eta_c-v$) for any $v$. Therefore, in the sensitivity analysis, we will vary the values of $\eta_n$ while fixing $\eta_c$ to 0 for convenience, and explore possible range of $v$ in the interpretation.

\section{Application to the Italian SHIW data}
\subsection{The Data} \label{sec:data}
The SHIW has been run every two years since 1965 with the only exception being that the 1997 survey was delayed to 1998. We denote by $t$ the generic survey year, and by $(t+1)$ the subsequent survey year. We define the target population as the set of households having at least one bank current account but neither debit cards nor credit cards at $t$. The treatment $Z$ is posed equal to 1 if, at $t+1$, the household (all members combined) possesses one and only one debit card and no credit cards, equal to 0 if, at $t+1$, the household possesses neither debit cards nor credit cards. The households with more than one debit card other than those possessing at least one credit card are excluded from the sample. The reason to exclude the households holding credit cards is that these households usually already possessed at least one debit card \citep{Mercatanti08}, and thus inclusion of these households would lead to imbalance in credit card holdings between treated and untreated households. Given that credit card is a payment instrument potentially affecting cash demand, this would overestimate the effect of debit cards. Therefore, a household for which $Z=1$ is characterized by having acquired their first (and only) debit card during the span $t\rightarrow (t+1)$. The two binary post-treatment variables $D$ identify the way in which the debit card is actually used: $D$ is posed equal to 1 if debit card is used to make ATM withdrawals at least one time per month on average (\textit{withdrawers}), or debit card is used to make payment at POS devices at least one time per month on average (\textit{POS users}). For the post-treatment variable \textit{withdrawers}, the relevant survey question asks the households how many withdrawals were made on average per month at ATM points during the survey's year. For \textit{POS users}, the relevant survey question asks the number of times, on average per month, the debit card was used directly at supermarket or shops to make payments by means of POS terminals. We conduct separate analysis with each of the two $D$ variables. The outcome $Y$ is the average cash inventory held by the household and is observed at $t+1$. The relevant survey question asks the sum of cash household usually have in the house to meet normal household needs.

The covariaties $X$ include the lagged outcome, some background demographic and social variables referred either to the household or to the head householder. The subset of covariates referred to the household includes the overall household income, wealth, the monthly average spending of the household on all consumer goods, and the following categorical variables: the number of earners, average age of the household, family size,  the Italian geographical macro-area where the household lives, the number of inhabitants of the town where the household lives. Those related to the head householder include age and education, both of which are categorical. As shown in \cite{Mercatanti14a}, the probability of a household having one debit card increases with income, the town size, education of the head householder, from the South to North of Italy, while decreases with the average age of the household.

Table \ref{tab:usage} shows, for the years 1995, 1998 and 2000 a non-negligible share of debit card holders who rarely use the card to withdraw cash from ATMs or to pay for purchases at POS devices at retailers. The share of non-users is less for withdraws at ATMs than for POS payments.

\begin{table}[ht!]
\renewcommand{\baselinestretch}{1.1}
\caption{Per cent of households with bank account, possessing one debit card and no credit cards, by debit card usage.}
\label{tab:usage}
\begin{center}
\begin{tabular}{lccc}
\hline
{ Year} & {Sample size} & {Less than one ATM withdrawal} & {Less than one POS payment}\\
        & & {per month on average} & {per month on average}\\
\hline
{ 1995} & {1727} & {23.2} & {87.2} \\
{ 1998} & {1645} & {16.8} & {68.7} \\
{ 2000} & {1857} & {19.3} & {59.2} \\
\hline
\end{tabular}
\end{center}
\end{table}

A simple descriptive cross-sectional analysis on the subsample of households observed in a single sweep of the survey shows the difference in average cash inventory between households possessing one debit card and households without a debit card is -121.6, -118.1, and -169.1 thousands of Italian Liras in 1995, 1998, and 2000, respectively. Though not sufficient to establish causal effects of debit cards on cash holding, this shows that consumers in Italy who possess debit cards hold less cash compared to those who do not on average.

SHIW is a repeated cross-section with a panel component, namely only a part of the sample comprises households that were interviewed in previous surveys. Our analysis will focus on the households observed for two consecutive surveys. Table \ref{tab:samplesize} reports the samples sizes for each span, $t\rightarrow (t+1)$, where $t=1993,95,98$, respectively by treated and untreated units. The relative frequency of untreated units alongside the total sample size has a considerable drop after 2000. Accordingly, the analysis will be limited to the span until 1998-00, the latest span with considerable share of both untreated units and total sample size.

\begin{table}[ht!]
\renewcommand{\baselinestretch}{1.1}
\caption{ Sample sizes and relative frequency of treated ($Z_i=1$) and untreated ($Z_i=0$) units for each span.}
\label{tab:samplesize}
\begin{center}
\begin{tabular}{lccccc}
\hline
${t\rightarrow (t+1)}$ & \multicolumn{2}{c}{$Z_i=1$} & \multicolumn{2}{c}{$Z_i=0$} & {Total}\\
\cline{2-5}
& { size} & { rel. freq.} & { size} & { rel. freq.} &  \\
\hline
1993-95 & { 164} & { .177} &{ 764} & { .823} & { 928} \\
1995-98 & { 143} & { .274} &{ 379} & { .726} & { 522} \\
1998-00 & { 114} & { .182} &{ 513} & { .818} & { 627} \\
\hline
\end{tabular}
\end{center}
\end{table}

\subsection{Assessment of covariate overlap and balance}

\begin{figure}[ht!]
\caption{Histograms of the estimated propensity score for the treated (blue) and the untreated (red). The first is for the span 1993 to 1995, the second is for the span 1995 to 1998, and the third is for the span 1998 to 2000.}
\label{fig:hist_ps}
\centering
\includegraphics[width=5cm,height=5cm]{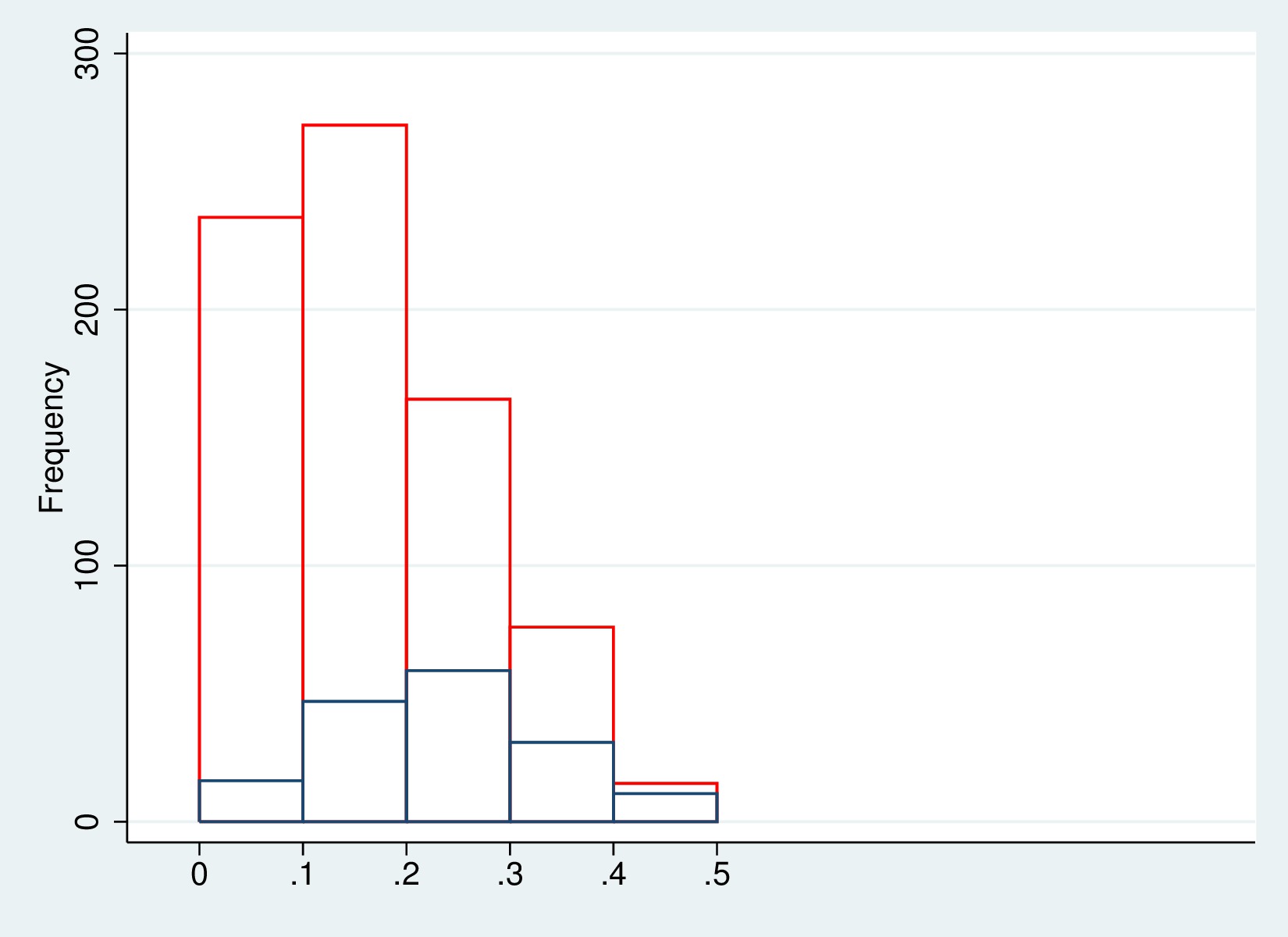}
\includegraphics[width=5cm,height=5cm]{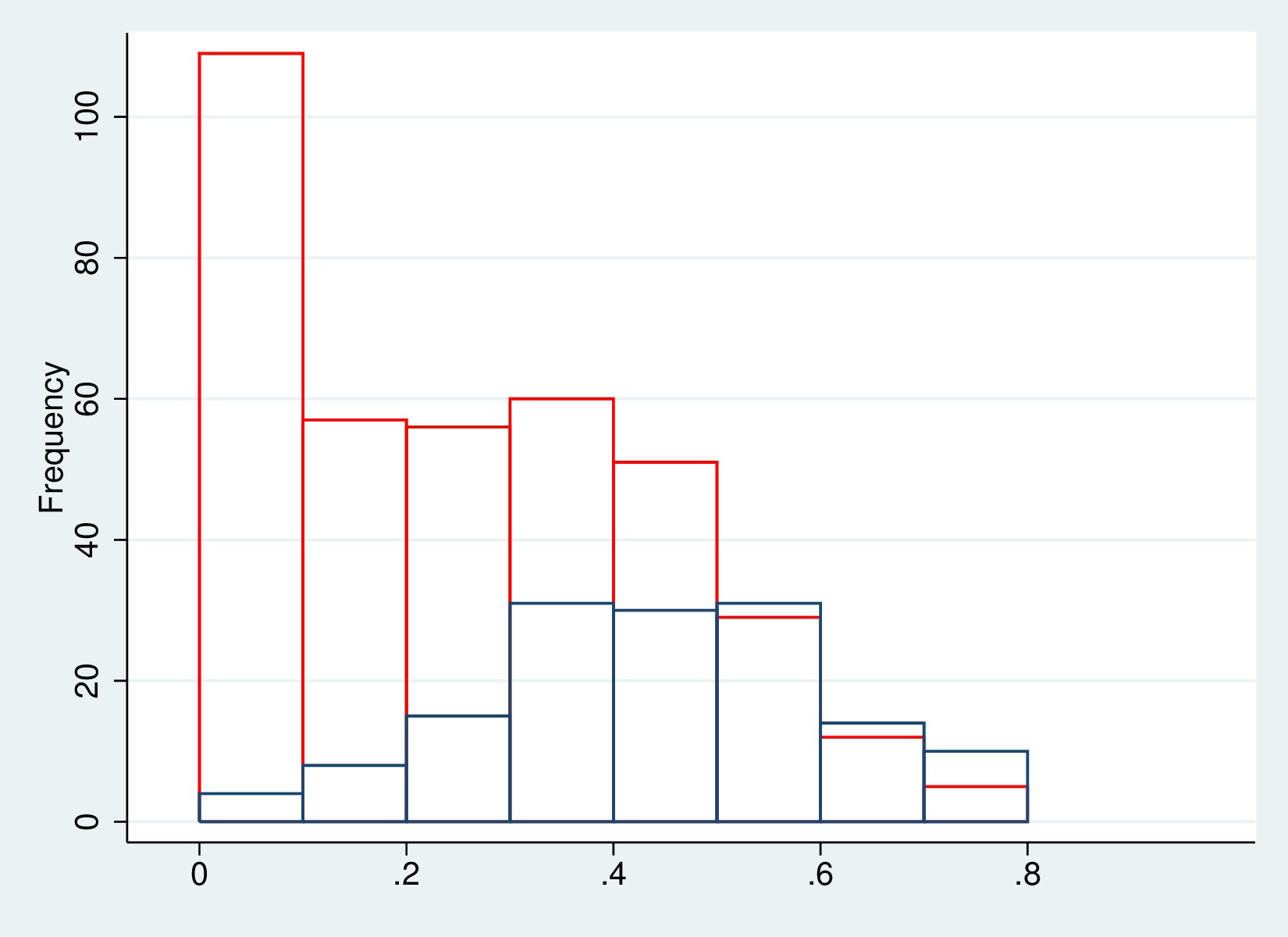}
\includegraphics[width=5cm,height=5cm]{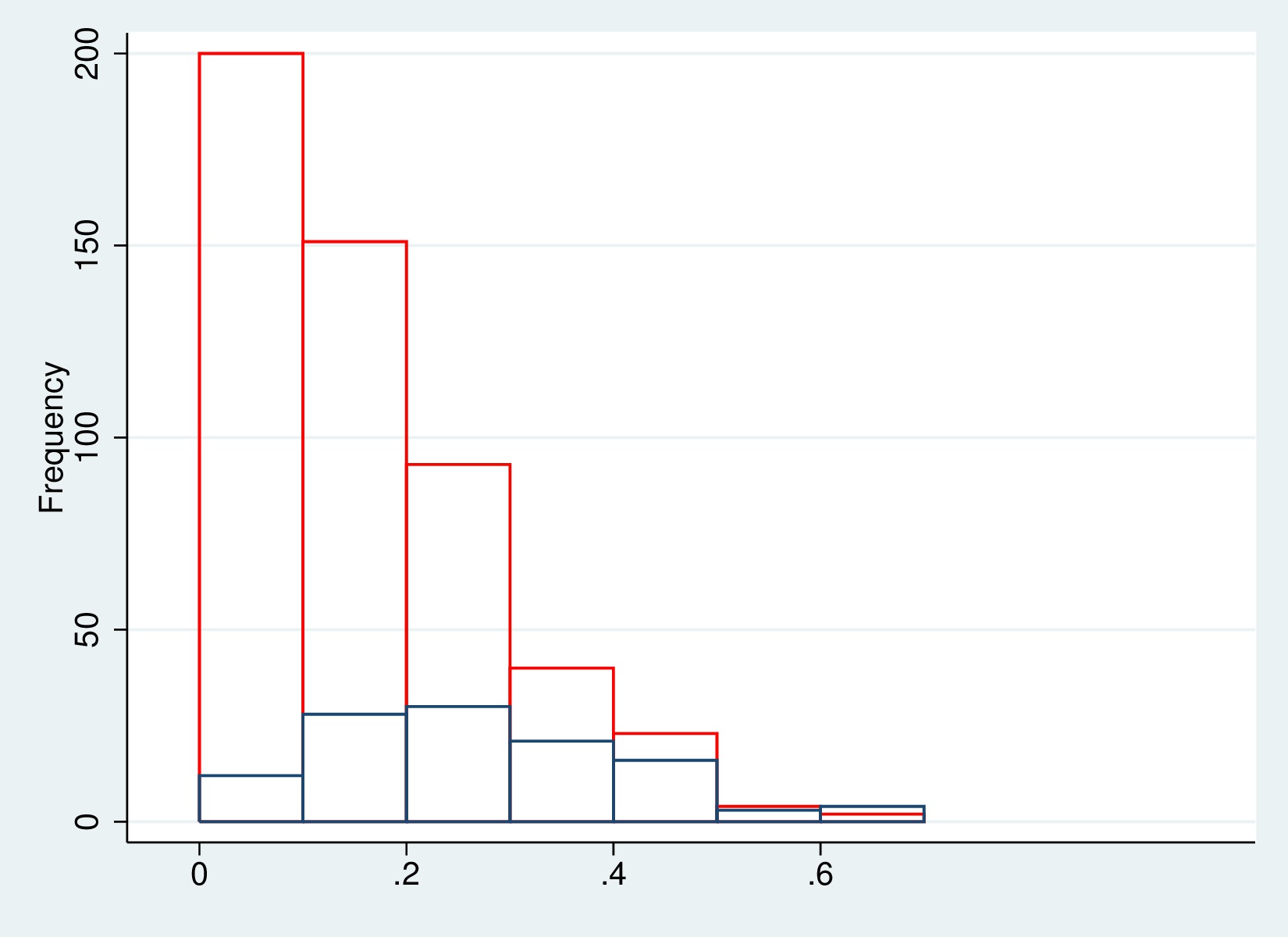}
\end{figure}

We first assess covariate overlap and balance in the studied sample. Figure \ref{fig:hist_ps} presents the histograms of estimated propensity scores for treated and untreated groups for each span, which were estimated from a logistic model with main effects of each covariate. The histograms show a satisfactory overlap in the support of the propensity score for all three spans, so that no trimming is needed, and further, this provides basis for our propensity score regression method based on the suggestions of \cite{Hade14}.

\begin{figure}[h]
\caption{Boxplots of the absolute standardized difference of all covariates.} \label{fig:boxplot}
\centering
\includegraphics[width=10cm,height=5cm]{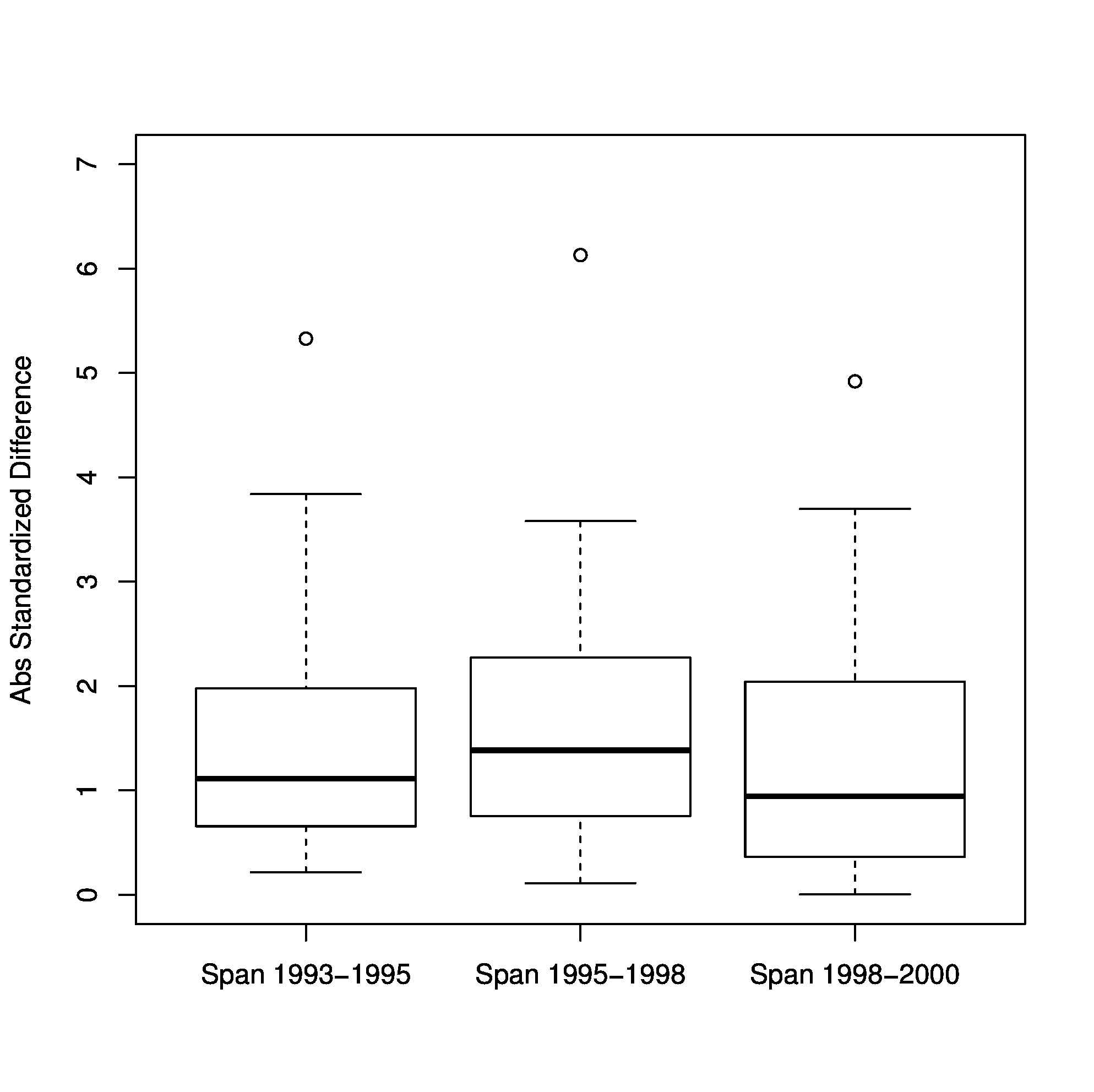}
\end{figure}

Figure \ref{fig:boxplot} reports, for each span, the boxplots of the absolute standardized differences (ASD) in covariates between treated and untreated group:
\begin{equation}
\text{ASD} = {\left|\frac{\sum_{i=1}^N X_i Z_i}{\sum_{i=1}^N Z_i} - \frac{\sum_{i=1}^N X_i (1- Z_i)}{\sum_{i=1}^N (1-Z_i)}\right|}\Bigg /{\sqrt{s_{1}^2/N_1 + s_{0}^2/N_0}},\label{eq:ASD}
\end{equation}
where $N_z$ is the number of units and $s_z^2$ is the standard deviation of the covariates in group $Z=z$ for $z=0,1$. The boxplots reveal significant imbalance in a large number of covariates between groups. Therefore, adjustment of covariate imbalance is crucial in this application.


\subsection{Results}
We fit the models in Section \ref{sec:models} via the EM algorithm. Table \ref{tab:treatednonusers} compares the proportion of never-users among the treated units estimated from the model with and without the sensitivity parameter $\xi$, calculated as follows:
$$\Pr(S=n|Z=1)=\int_{e(x)}\Pr(S=n|Z=1,~e(x))\cdot \Pr(e(x)|Z=1)~dx,$$
where $\Pr(S=n|Z=1,~e(x))$ is the model for $S$ and $\Pr(e(x)|Z=1)$ is approximated by the observed distributions of the estimated $e(x)$ among the treated units. As a reference, we also present the moment estimates of this quantity, which is the proportion of non-users in the group of households with one debit card. 
Table \ref{tab:treatednonusers} shows that the moment estimates closely resemble estimates from the model with the sensitivity parameter $\xi$ (model \eqref{eq:smodel_confound}) in all spans, but differ much from those from the model without the parameter (model \eqref{eq:smodel}), suggesting the latter is subject to large bias due to unmeasured confounding (elaborated later).

Table \ref{tab:sens_MLE} reports the results obtained from the sensitivity model with $\xi$ free in equation \eqref{eq:smodel_confound}, and with $\eta_c=0$ and varying values of $\eta_n$ in equation \eqref{eq:ymodel_confound}. The parameter $\eta_n$ represents the level of $Y$-confounding for never-users by the difference in average cash inventories between never-users with one debit card and without debit card. Absolute values of $\eta_n$ greater than 400 thousands of Italian Lira (LIT hereafter) is considered too high given the reported values of average cash inventories, therefore we limit the range of examined values for $\eta_n$ between -400 and 400 thousands LIT. The average cash inventories for never-users with one debit card is observable and is in the ranges from 630 to 788, and from 703 to 990 thousands of LIT  for POS users and withdrawers, respectively. Estimates of $\xi$ are negative under each scenario. This suggests that, conditionally on the propensity score, the probability of being non-users is lower among treated units (card-holding households) than untreated units (non card-holding households), revealing a violation of the unconfoundedness assumption due to $S$-confounding. Indeed we found the untreated compliers most consist of a small group of households with high cash inventories.  The span 1995-1998 emerges with the highest estimated value of $\alpha_0$ (the intercept of the model for $S$), implying a high proportion of non-users in the group of untreated households. Estimates of the CATEs and CATTs are consistently negative with small standard errors. The span 1995-1998 shows larger values of the two estimated causal effects, likely due to the detected higher $S$-confounding. The estimated values of $\alpha_0$, $\alpha$, $\xi$ and in particular of our targeted estimand, CATT, remain stable across the range of $\eta_c\neq0$,  while the estimated CATEs are equal to those obtained under $\eta_c=0$ minus the alternative value of $\eta_c$. Table \ref{tab:sens_MLE} shows that, for each of the six considered scenario, the estimated CATT is insensitive to the degree of unmeasured confounding within a reasonable range. The estimated CATTs for withdrawers are comparatively more stable, in particular the span 1998-2000 that shows values contained in the small range from -1547 to -1572 thousands LIT. The estimated CATTs for POS users are more sensitive to the proposed values of $\eta_n$ presumably because of the larger shares of non-users in comparison to withdrawers (Table \ref{tab:usage}). However, these values remain within plausible ranges, with the only exception of the CATT corresponding to span 1993-1995 and $\eta_n=-400$ that is appreciably different to the CATTs obtained for the alternative values of $\eta_n$. The span 1998-2000 shows again stabler results: -1573 to -1687 thousands LIT, presumably due to the smaller share of POS non-users in comparison to the other two spans.

To better understand the scale of these results, Table $\ref{tab:AOTC}$ reports the percentage reduction in the cash inventories due to the use of the debit card, $\frac{\mbox{CATT}}{\mbox{AOTC}-\mbox{CATT}}$, where AOTC is the Average Outcome for the Treated Compliers. For 1993-95 and 1998-00, the ratios are greater for POS users than for withdrawers. This is reasonable because individuals who used debit cards to pay at POS usually also used cards to withdraw cash at ATM. For 1995-98 the difference between POS users and withdrawers decrease in that people usually start to use debit cards to withdraw and subsequently to pay at POS, so that the longer the span the more likely that the groups of POS users and withdrawers coincide. Overall the estimated reduction in household cash inventory due to the use of the debit card is remarkable, ranging between 78\% and 81\% for the span 1995-1998, between 75\% and 78\% for the POS users and between 67\% and 73\% for the withdrawers for the other two spans.


 \begin{table}[ht]
\caption{Estimated proportion of never-users in the group of treated, $\Pr(S=n|Z=1)$, from the model with the sensitivity parameter $\xi$ in comparison to the model without sensitivity parameters and moment estimates.}
\label{tab:treatednonusers}
\begin{center}
\begin{tabular}{llccc}
\hline
&&  1993-1995 & 1995-1998 & 1998-2000 \\
\hline
POS users& Model with sensitivity parameter $\xi$  & .933 & .780 & .690   \\
& Model w/o sensitivity parameters  & .886 & .308 & .612   \\
& Moment Estimate & .933 & .783 & .690\\
\hline
Withdrawers & Model with sensitivity parameter $\xi$  & .301 & .297 & .140   \\
& Model w/o sensitivity parameters   & .700 & .635 & .407   \\
& Moment Estimate  & .299 & .300 & .140 \\
\hline
\end{tabular}
\end{center}
\end{table}

\begin{table}[ht]
\caption{MLE of $\alpha _{0}$, $\alpha$, $\xi$, $\theta$ (CATE) and CATT when $\eta_{c}=0$ and for fixed values of $\eta _{n}$. CATE and CATT denominated in thousands of Italian Lira. Standard errors are in the parenthesis.}
\label{tab:sens_MLE}
\begin{center}
\begin{footnotesize}
\begin{tabular}{llrrrrr}
\hline
    &$\eta _{n}$         & -400           & -200           & 0              & 200            & 400            \\
\hline
&$\hat{\alpha}_{0}$ & 3.38 (.43)  & 2.32 (.27) & 2.34 (.26) & 2.42 (.25) & 2.48 (.25) \\
93-95  &$\hat{\alpha}$ & .35 (2.46) & -1.52 (1.40) & -2.00 (1.28) & -2.86 (1.21) & -3.52 (1.17) \\
POS users &$\hat{\xi}$ & -.84 (.48) & -.67 (.36) & -.77 (.35) & -.91 (.35) & -1.03 (.35) \\
& CATE & -2401.9 (291.4) & -1708.6 (168.4) & -1673.5 (158.2) & -1617.5 (156.2) & -1574.9 (161.7) \\
& CATT  & -2429.6 (911.8) & -1730.6 (460.5) & -1698.1 (455.9) & -1647.9 (450.1) & -1609.0 (471.8)  \\
\hline
&$\hat{\alpha}_{0}$ & 2.39 (.24) & 2.38 (.23) & 2.37 (.23) & 2.37 (.23) & 2.38 (.23) \\
93-95 & $\hat{\alpha}$ & -1.84 (1.11) & -1.97 (1.08) & -2.06 (1.07) & -2.14 (1.06) & -2.22 (1.07) \\
withdrawers &$\hat{\xi}$ & -2.80 (.24) & -2.76 (.23) & -2.74 (.22) & -2.72 (.22) & -2.71 (.22) \\
&CATE & -1506.4 (58.7) & -1493.6 (55.2) & -1483.2 (53.4) & -1484.3 (52.9) & -1469.2 (53.4) \\
&CATT  & -1562.7 (229.5) & -1551.3 (229.1) & -1542.8 (216.9) & -1536.4 (220.4) & -1531.0 (217.9) \\
\hline
&$\hat{\alpha}_{0}$ & 3.26 (.48) & 3.19 (.46) & 3.16 (.46) & 3.12 (.45) & 3.11 (.45) \\
95-98  & $\hat{\alpha}$ & 2.23 (1.13) & 1.79 (1.07) & 1.64 (1.05) & 1.31 (1.02) & 1.11 (.99) \\
POS users &$\hat{\xi}$ & -2.94 (.48) & -2.68 (.43) & -2.59 (.41) & -2.42 (.39) & -2.33 (.39) \\
&CATE & -2881.9 (207.2) & -2767.1 (182.6) & -2720.6 (174.6) & -2637.0 (170.6) & -2596.3 (174.3) \\
&CATT  & -2902.8 (479.5) & -2805.6 (512.6) & -2763.2 (477.2) & -2691.6 (432.8) & -2658.3 (493.8) \\
\hline
&$\hat{\alpha}_{0}$ & 3.67 (.45) & 3.65 (.44) & 3.62 (.44) & 3.59 (.43) & 3.57 (.43) \\
95-98  &$\hat{\alpha}$ & -.26 (.96) & -.31 (.94) & -.37 (.93) & -.49 (.92) & -.62 (.91) \\
withdrawers &$\hat{\xi}$ & -4.41 (.43) & -4.37 (.42) & -4.32 (.41) & -4.23 (.40) & -4.15 (.39) \\
&CATE & -2775.4 (169.4) & -2744.1 (159.5) & -2706.7 (152.8) & -2646.3 (147.3) & -2592.1 (144.4) \\
&CATT & -2739.6 (305.0) & -2710.0 (423.9) & -2674.9 (426.8) & -2619.3 (433.0) & -2571.0 (403.4) \\
\hline
&$\hat{\alpha}_{0}$ & 2.35 (.23) & 2.25 (.22) & 2.19 (.22) & 2.15 (.22) & 2.11 (.22) \\
98-00 & $\hat{\alpha}$ & -1.89 (.93) & -1.75(.91) & -1.69(.90) & -1.70 (.89) & -1.87 (.88) \\
POS users &$\hat{\xi}$ & -1.01 (.30) & -.94 (.29) & -.90 (.28) & -.85 (.29) & -.77 (.29) \\
&CATE  & -1785.4 (137.5) & -1761.4 (126.0) & -1739.1 (121.2) & -1709.9 (121.0) & -1652.8 (63.6) \\
&CATT  & -1687.3 (391.0) & -1655.1 (379.4) & -1632.3 (375.5) & -1608.3 (373.4) & -1573.0 (366.0) \\
\hline
&$\hat{\alpha}_{0}$ & 2.00 (.24) & 1.98 (.24) & 1.97 (.23) & 1.96 (.24) & 1.95 (.23) \\
98-00 & $\hat{\alpha}$ & -.45 (1.02) & -.42 (1.02) & -.41 (1.02) & -.41 (1.02) & -.43 (1.03) \\
withdrawers &$\hat{\xi}$ & -3.69 (.34) & -3.68 (.33) & -3.67 (.32) & -3.66 (.32) & -3.65 (.33) \\
&CATE & -1643.9 (72.6) & -1635.6 (70.9) & -1628.8 (69.7) & -1623.4 (69.5) & -1619.4 (69.9) \\
&CATT & -1572.0 (244.6) & -1563.1 (248.1) & -1555.6 (239.6) & -1550.9 (246.3) & -1547.8 (244.2) \\
\hline
\end{tabular}
\end{footnotesize}
\end{center}
\end{table}

\begin{table}[ht]
\caption{Percentage reduction in the cash inventories due to the use of the debit card, $\frac{\mbox{CATT}}{\mbox{AOTC}-\mbox{CATT}}$, across a range of $\eta _{n}$ values.}
\label{tab:AOTC}
\begin{center}
\begin{tabular}{lccccc}
\hline
           & $\eta _{n}=-400$ & $\eta _{n}=-200$ & $\eta _{n}=0$ & $\eta _{n}=200$ &$\eta _{n}=400$            \\
\hline
93-95: POS users   & -.832 & -.779 & -.776 & -.770 & -.766 \\
93-95: withdrawers & -.678 & -.676 & -.675 & -.674 & -.673 \\
\hline
95-98: POS users   & -.813 & -.808 & -.806 & -.802 & -.797 \\
95-98: withdrawers & -.793 & -.791 & -.789 & -.785 & -.782 \\
\hline
98-00: POS users   & -.758 & -.755 & -.752 & -.749 & -.745 \\
98-00: withdrawers & -.730 & -.729 & -.728 & -.728 & -.727 \\
\hline
\end{tabular}
\end{center}
\end{table}

\section{Discussion}
In this paper we quantify the causal effect of the use of debit cards on households cash inventories in Italy. A principal stratification model integrated with sensitivity parameters allows to simultaneously account for issues including the non-negligible share of households who hold one debit card but do not use it, the questionable definition of the use of cards as a treatment under the potential outcome approach, and possible violation of the unconfoundedness assumption.

Our results suggest considerable causal effects: the reduction on cash inventories for households who use the debit card is consistently between 70 and 80 per cent during 1993-2000. We have evaluated short-term effects here with only one to one and a half years long study period on average. In fact, the SHIW data does not provide information about the moment a household has acquired its debit cards; we only know it has happened during the two, or three, years of the considered span. Therefore the high estimated effects on cash holding also signal that the use of cards quickly affects the reduction of the amount of cash held at home.

Via the sensitivity analysis we have identified a high level of unmeasured confounding that otherwise would have biased the results. Source for the confounding primarily lies in the part of compliers who have high level of cash inventories even without possessing debit cards. Indeed \cite{Mercatanti14a} also shows debit cards holders generally have higher level of income, wealth and education of the householder in comparison to households without debit cards. Therefore it is plausible that compliers are in high social and economic statuses. Quantification of the causal effects for a larger sub-population of compliers that include households in a broad range of social classes, could be achieved by extending the observed period, which would reduce the proportion of never-users. In fact, it is plausible that less reactive households would start to use the cards over time. However, this analysis is not feasible with SHIW data for two reasons: first, extended period of observations would greatly reduce the sample sizes; second, given the increased diffusion of debit cards, the control group size would collapse over time. Nonetheless, one could still apply the same causal model to suitable datasets that allow for extending the temporal effects of debit cards while maintaining an adequate sample size for the untreated group.

Our sensitivity model for the outcome is not identifiable from a frequentist perspective, with only the sum of the two but not individual sensitivity parameters identifiable. From a Bayesian perspective, the model is weakly identifiable given proper prior distributions for the parameters. When available, a secondary outcome variable modeled jointly with the primary outcome would sharpen the analysis \citep{MealliPacini13, Mercatanti14b} regardless of the mode of inference. Searching for a suitable auxiliary outcome is a direction of our future research.

\bibliographystyle{natbib}
\bibliography{cashdemand}

\end{document}